\documentclass[12pt]{article}
\usepackage{amsfonts,amssymb,epsfig}
\begin{document}

\newcommand{\be}{\begin{equation}}
\newcommand{\ee}{\end{equation}}
\newcommand{\bea}{\begin{eqnarray}}
\newcommand{\eea}{\end{eqnarray}}
\newcommand{\beas}{\begin{eqnarray*}}
\newcommand{\eeas}{\end{eqnarray*}}

\baselineskip 14 pt
\parskip 12 pt

\begin{titlepage}
\begin{flushright}
{\small BROWN-HET-1339} \\
{\small CU-TP-1080} \\
{\small hep-th/0212246}
\end{flushright}

\begin{center}

\vspace{2mm}

{\Large \bf Quasiparticle picture of black holes and \break the entropy--area relation}

\vspace{3mm}

Norihiro Iizuka${}^1$, Daniel Kabat${}^1$, Gilad Lifschytz${}^2$ and David A.\ Lowe${}^3$

\vspace{1mm}

${}^1${\small \sl Department of Physics} \\
{\small \sl Columbia University, New York, NY 10027} \\
{\small \tt iizuka, kabat@phys.columbia.edu}
\vspace{1mm}

${}^2${\small \sl Department of Mathematics and Physics and CCMSC} \\
{\small \sl University of Haifa at Oranim, Tivon 36006, Israel} \\
{\small \tt giladl@research.haifa.ac.il}

${}^3${\small \sl Department of Physics} \\
{\small \sl Brown University, Providence, RI 02912} \\
{\small \tt lowe@het.brown.edu}

\end{center}

\vskip 0.3 cm

\noindent
We propose an effective description of 0-brane black holes, in which
the black hole is modeled as a gas of non-interacting quasi-particles
in the dual quantum mechanics.  This simple model is shown to account
for many of the static thermodynamic properties of the black hole.  It
also accounts for dynamical properties, such as the rate at which
energy gets thermalized by the black hole.  We use the model to show
that the entropy of the quantum mechanics is proportional to the black
hole horizon area in Planck units.

\end{titlepage}

%%%%%%%%%%%%%%%%%%%%%%%%%%%%%%%%%%%%%%%%%%%%%%%%%%%%%%%%%%%%%%%%%%%%%%%%%
\section{Introduction}
%%%%%%%%%%%%%%%%%%%%%%%%%%%%%%%%%%%%%%%%%%%%%%%%%%%%%%%%%%%%%%%%%%%%%%%%%

The Hawking-Bekenstein relationship \cite{HawkingBekenstein} between
entropy and area is a fundamental property of quantum gravity.  In
recent years it has become possible to build microscopic models of
black holes, and show that the entropy -- area relationship indeed
holds \cite{StromingerVafa,Maldacena:1996gb,Breckenridge:1996is,Horowitz:1996ay,Breckenridge:1996sn,Horowitz:1996ac}.
But in these calculations the underlying reason for a universal
connection between entropy and area remains obscure, since the black
hole microstates are only explicitly constructed in a limit where the
geometry is singular.

The advent of the AdS/CFT correspondence \cite{maldacena} and its
generalizations \cite{imsy} should make it possible to better
understand the relationship between entropy and area.  In the AdS/CFT
framework the entropy of the black hole can be identified with the
thermal entropy of the CFT, while the area of the horizon should in
principle be calculable within the CFT.  However to date not much
progress has been made in this direction.  The problem is that we do
not understand in detail how quasi-local bulk gravitational physics
emerges from the dual CFT.  This makes the direct CFT definition of
horizon area difficult.

We will work in the context of the duality between 0-brane quantum
mechanics and 0-brane black holes \cite{imsy}.  In the regime where
the black holes are well-described by supergravity the quantum
mechanics is strongly coupled.  In \cite{kl,kll,ikll} we developed a
mean-field approximation scheme for the strongly-coupled quantum
mechanics, and showed that it captures some of the essential physics
of the dual gravity theory.  Similar approximations have been used in
\cite{IIB} in the framework of the IKKT matrix model \cite{ikkt}.

In the end the mean-field approximation suggests the following very
simple picture of the quantum mechanics.  The finite-temperature
quantum mechanics has an effective description in terms of a
collection of non-interacting quasi-particles \footnote{The
quasi-particles we consider have nontrivial two-point functions, but
no multi-particle interactions.}, which we identify with the
individual matrix elements appearing in the matrix fields $X(t)$.
There are a total of $N^2$ quasi-particles.  Some of the
quasi-particles have an energy of order the temperature and can be
thermally excited.  The remaining quasi-particles are too heavy to be
thermally excited.  The light quasi-particles are unstable, and decay
with a characteristic lifetime of order the inverse temperature.

We would like to stress that the validity of the quasi-particle
picture we have described is compatible with, but does not rely on,
the validity of any particular mean-field approximation.  But given
our mean-field results, the quasi-particle properties can be extracted
as follows.  The imaginary time two-point function $\langle X(\tau)
X(0) \rangle$ has a spectral representation, which we study in
ref.~\cite{ikll} and appendix A of the present work.  There we find
that the spectral density $\rho(\omega)$ has two peaks.  One peak is
concentrated at a frequency of order the temperature and the other is
concentrated at a frequency of order one in 't Hooft units
\cite{ikll}.  Moreover in appendix A we show that the low-frequency
peak has a width of order the temperature, which implies that the
lifetime of the light quasi-particles is approximately $1/T$.  There
are a total of $N^2$ quasi-particles.  But at temperature $T$, only
\[
N_{\rm eff} = N^2 \int_{\hbox{\rm first peak}} d\omega \rho(\omega)
\]
of the quasi-particles can be thermally excited.  The number of light
quasi-particles depends on the temperature.  In \cite{kll} we computed
the entropy of the quantum mechanics in the mean-field approximation
and found agreement with the black hole entropy up to factors of order
unity.  Thus we take the number of light quasi-particles to be about
equal to the entropy of the black hole, $N_{\rm eff} \approx S_{\rm
BH}$ up to constant factors.

To summarize, our picture of the quantum mechanics is that it consists
of $N_{\rm eff} \approx S_{\rm BH}$ light quasi-particles, each with
an energy $\sim T$ and a lifetime $\sim 1/T$.  We propose that these
light quasi-particles give a holographic description of the dual black
hole.  That is, in the dual gravitational description we take the
quasi-particles to correspond to the stretched horizon degrees of
freedom, which provide a holographic description of the black hole
together with its low-energy excitations.  We take the stretched
horizon to be located where the proper temperature is equal to the
Planck temperature.  As we will see in section \ref{thermalization},
this is necessary to have a consistent stretched horizon description
of the low-energy excitations of the black hole.

We should point out that the picture of black holes we are advocating
draws heavily on previous work.  It is closely related to the membrane
paradigm description of black holes \cite{membrane}.  It also has
significant overlap with the work of Susskind and others on black hole
complementarity \cite{stu,s2,StHW,KLMO,Lowe:1995pu,Lowe:1995ac,Lowe:1999pk}.

An outline of this paper is as follows.  In section 2 we show that the
quasi-particle picture correctly reproduces the equilibrium
thermodynamic properties of the black hole, including a relationship
between the entropy and radius of the black hole.  In section 3 we
compute the thermalization rate of the black hole, and use this to
derive the relationship between entropy and horizon area. Section 4
contains some comments and conclusions. In appendix A we relate the
quasi-particle picture of 0-brane quantum mechanics to our previous
results on the mean-field approximation and deduce the quasi-particle
lifetime.  In appendix B we study quasi-normal modes of a scalar field
in the black hole background, and show that their lifetime matches
that of the quasi-particles.

%%%%%%%%%%%%%%%%%%%%%%%%%%%%%%%%%%%%%%%%%%%%%%%%%%%%%%%%%%%%%%%%%%%%%%%%%
\section{Equilibrium properties}
\label{equilibrium}
%%%%%%%%%%%%%%%%%%%%%%%%%%%%%%%%%%%%%%%%%%%%%%%%%%%%%%%%%%%%%%%%%%%%%%%%%

In this section we study four equilibrium thermodynamic properties of
the quasi-particle gas, namely the entropy, energy, specific heat and
radius.  We will argue that they have the same qualitative behavior on
both sides of the duality.

First let's consider the quantum mechanics side.  We model the quantum
mechanics as made up of $N_{\rm eff}$ harmonic oscillators with
frequency $\omega \sim T$.  Thus at temperature $T$ we have an entropy
\begin{equation}
S_{\rm QM} = \Bigl(1 - \beta {\partial \over \partial \beta}\Bigr) \log Z
\approx N_{\rm eff} \,,
\label{entro}
\end{equation}
an energy
\[
E_{\rm QM} = - {\partial \over \partial \beta} \log Z \approx N_{\rm eff} T \,,
\]
and a specific heat
\[
c_{\rm QM} = \beta^2 {\partial^2 \over \partial \beta^2} \log Z \approx N_{\rm eff} \,.
\]
On the black hole side the entropy is given by \cite{imsy}
\[
S_{\rm BH} \sim N^2 (T / \lambda^{1/3})^{9/5}
\]
where $\lambda = g^2_{\rm YM} N$ is the 't Hooft coupling of the
quantum mechanics.  Thus the entropies agree provided the number of
quasi-particles is about the same as the entropy of the black hole,
$N_{\rm eff} \approx S_{\rm BH}$.  At the moment we must simply assume
that this relationship holds, although up to factors of order one, it
can be shown to follow from the mean-field approximation \cite{kll}.
The energy of the black hole is
\[
E_{\rm BH} \sim N^2 \lambda^{1/3} (T / \lambda^{1/3})^{14/5} \approx T S_{\rm BH}
\]
in agreement with the quasi-particle prediction.  Finally the specific
heat of the black hole is
\[
c_{\rm BH} \sim N^2 (T/\lambda^{1/3})^{9/5} \approx S_{\rm BH}
\]
again in agreement with the quasi-particle prediction.

Now let us consider the size of the black hole in the quantum
mechanics.  The matrix fields $X(t)$ are related to a collection of
canonically normalized harmonic oscillators by $X = ({\lambda \over
N})^{1/2} \,\, Y$.  So we identify
\be
\label{radius}
R_{\rm h}^2 = {1 \over N} \langle {\rm Tr} X^2 \rangle
= {\lambda \over N^2} \langle {\rm Tr} Y^2 \rangle
\ee
with the squared radius of the black hole.  Following Susskind
\cite{Lenny} we suppress high-frequency contributions to the
expectation value, so that only the $N_{\rm eff}$ light degrees of
freedom contribute to the trace in (\ref{radius}).  This procedure is
made possible by the clear separation of scales in the double-peaked
spectral density.  Thus
\[
R_{\rm h}^2 \approx {\lambda \over N^2} N_{\rm eff} \langle x^2 \rangle
\approx {\lambda N_{\rm eff} \over N^2 T}
\]
where we have used the fact that at temperature $T$ a single harmonic
oscillator with $\omega \approx T$ has $\langle x^2 \rangle \approx
1/T$.  That is, the quantum mechanics predicts a relation between the
horizon radius and the entropy \cite{ikll}
\begin{equation}
R_{\rm h}^2 \approx {\lambda S \over N^2 T}\,.
\label{redrel}
\end{equation}
Rather remarkably, this relationship indeed holds for 0-brane black
holes \cite{kl}, since the supergravity expression for the horizon
radius is\footnote{denoted $U_0$ in \cite{imsy}}
\[
R_{\rm h} \sim \lambda^{1/3} \Bigr(T / \lambda^{1/3}\Bigl)^{2/5}\,.
\]

%%%%%%%%%%%%%%%%%%%%%%%%%%%%%%%%%%%%%%%%%%%%%%%%%%%%%%%%%%%%%%%%%%%%%%%%%
\section{Horizon area and thermalization time}
%%%%%%%%%%%%%%%%%%%%%%%%%%%%%%%%%%%%%%%%%%%%%%%%%%%%%%%%%%%%%%%%%%%%%%%%%

%%%%%%%%%%%%%%%%%%%%%%%%%%%%%%%%%%%%%%%%%%%%%%%%%%%%%%%%%%%%%%%%%%%%%%%%%
\subsection{Measuring the area}
%%%%%%%%%%%%%%%%%%%%%%%%%%%%%%%%%%%%%%%%%%%%%%%%%%%%%%%%%%%%%%%%%%%%%%%%%

Our original motivation for this paper was to gain insight into the
relationship between entropy and area.  In this section we report on
our progress in this direction.

The first step is to find a meaningful way of measuring the area of
the horizon.  For black holes in asymptotically flat space a simple
way to measure the horizon area is to perform a scattering experiment.
The classical absorption cross section (geometric optics cross
section) for high-energy particles is proportional to the area of the
black hole.  However this is not true for black holes in
asymptotically AdS-like spaces.  For example in string frame a null
geodesic with energy $E$ and angular momentum $\ell$ in the
near-horizon geometry (\ref{metric}) obeys
\[
\left({dU \over d\tau}\right)^2 + {\ell^2 U^5 \over c \lambda}
\left(1 - {U_0^7 \over U^7}\right) = E^2\,.
\]
Here $c = 240 \pi^5$, $\lambda$ is the 't Hooft coupling, and $U$ is a
radial coordinate with the horizon located at $U=U_0$.  All geodesics
fall into the horizon, and there are no classical scattering states.
Another possibility is to work at low energy.  For minimally coupled
scalars in asymptotically flat space the absorption cross section at
zero frequency is exactly the area of the black hole \cite{dgm}.  In
asymptotically AdS-like spaces one must regard scattering as a
tunneling problem and use non-normalizable solutions to the wave
equation to compute `absorption' cross-sections \cite{absorption},
which in the low-energy limit do turn out to be proportional to the
horizon area.  It would be interesting to apply this approach to
0-brane black holes.  Unfortunately by working in a low energy limit
one loses any intuitive connection between cross section and area.

We therefore turn to a different way of measuring horizon area, namely
the fact that a hot object with surface area $A$ will emit blackbody
radiation at a rate proportional to $A$.\footnote{A related procedure
(in a low-frequency limit) was used to define horizon area in
\cite{dmhkrs}, for black holes in asymptotically flat space.}
Following \cite{membrane,stu} we model the black hole as made up of a
set of degrees of freedom living at a stretched horizon which is
located just outside the true event horizon.  The Stefan-Boltzmann law
gives the rate at which the stretched horizon emits energy in outgoing
Hawking radiation.
\[
{d E^{\rm out}_{\rm proper} \over dt_{\rm proper}} \sim A T_{\rm proper}^{10}\,.
\]
Here $E_{\rm proper}$, $t_{\rm proper}$ and $T_{\rm proper}$ are
proper energies, times and temperatures measured at the stretched
horizon.  It is convenient to multiply this equation by $-g_{tt}$, to
get a relation between the corresponding Schwarzschild quantities.
\begin{equation}
\label{therma}
{d E^{\rm out} \over dt} = A T_{\rm proper}^8 T^2~.
\end{equation}
Again this is the rate at which the stretched horizon emits
energy.\footnote{For black holes in asymptotically flat space only a
tiny fraction of this energy flux ever reaches infinity, as most of
the outgoing radiation eventually falls back onto the horizon
\cite{membrane}.}  In thermal equilibrium, of course, this is balanced
against an equal and opposite flux of infalling energy.  From the
membrane paradigm point of view, the energy flux (\ref{therma}) is the
rate at which energy leaves the membrane, turns around, and eventually
falls back onto the membrane.  Thus in thermal equilibrium it measures
the rate at which energy gets redistributed among the quasi-particle
degrees of freedom which live on the stretched horizon.  We take the
stretched horizon to be located at a radius where the proper
temperature is equal to the Planck temperature.\footnote{For more on
this choice see section \ref{thermalization} and the discussion
section.}  The rate at which energy is redistributed on the stretched
horizon is then
\begin{equation}
\label{therma1}
{d E \over dt} = \frac{A}{\ell_P^8} T^2.
\end{equation}
We will use this to measure the horizon area in Planck units.

%%%%%%%%%%%%%%%%%%%%%%%%%%%%%%%%%%%%%%%%%%%%%%%%%%%%%%%%%%%%%%%%%%%%%
\subsection{Relating entropy to area}
%%%%%%%%%%%%%%%%%%%%%%%%%%%%%%%%%%%%%%%%%%%%%%%%%%%%%%%%%%%%%%%%%%%%%

The quantum mechanics has on average $N_{\rm eff}$ quasi-particles
each with an energy $\sim T$ and a lifetime $\sim 1/T$.  In thermal
equilibrium as these quasi-particles decay new quasi-particles are
continually created.  The rate at which the total energy gets
redistributed by this process is
\begin{equation}
{dE \over dt} \approx N_{\rm eff} T^2\,.
\end{equation}
Comparing this to the black hole result (\ref{therma1}) we are led to
identify
\begin{equation} 
N_{\rm eff} \sim \frac{A}{\ell_P^8}\,.
\end{equation}
But the entropy of the quantum mechanics $S_{\rm QM} \sim N_{\rm
eff}$.  Thus
\begin{equation}
\label{SA}
S_{\rm QM} \sim {A \over \ell_P^8}\,.
\end{equation}
This gives a simple direct connection between the area of the black
hole as measured in Planck units and the entropy of the dual quantum
mechanics.  Up to a numerical coefficient of order one it agrees with
the Hawking--Bekenstein formula.

%%%%%%%%%%%%%%%%%%%%%%%%%%%%%%%%%%%%%%%%%%%%%%%%%%%%%%%%%%%%%%%%%%
\subsection{Thermalization time}
\label{thermalization}
%%%%%%%%%%%%%%%%%%%%%%%%%%%%%%%%%%%%%%%%%%%%%%%%%%%%%%%%%%%%%%%%%%

We conclude by computing the thermalization time of the quasi-particle
gas, and showing that it matches the thermalization time of the black
hole.

First let us consider the black hole side.  In thermal equilibrium the
rate at which energy is radiated by the stretched horizon
(\ref{therma}) is balanced against an equal and opposite flux of
infalling energy.  But suppose we perturb the temperature of the black
hole $T \rightarrow T + \Delta T$, while keeping the temperature
outside the stretched horizon fixed.  Then there will be a net flux of
energy out of the black hole, given by
\[
{d \Delta E \over dt} \sim A T_{\rm proper}^8 T \Delta T\,.
\]
But the perturbation to the energy is $\Delta E = c \Delta T$, where
$c$ is the specific heat, so
\[
{d \Delta E \over dt} \sim {1 \over c} \, A T_{\rm proper}^8 T \Delta E\,.
\]
Recall that for these black holes the specific heat is proportional to
the entropy, $c \sim S \sim A T_{\rm proper}^8$.  Thus the
thermalization time of the black hole is
\be
\label{therma2}
\tau_{\rm BH} \sim 1/T \,.
\ee
Now consider the quantum mechanics side.  The thermalization time of
the quantum mechanics is set by the quasi-particle lifetime, so
\[
\tau_{\rm QM} \sim 1/T
\]
in agreement with the black hole.

Note that the result (\ref{therma2}) relies on putting the stretched
horizon at the Planck temperature.  In appendix B we show that
(\ref{therma2}) matches the lifetime of quasi-normal excitations of
the black hole background.  This matching is expected if the stretched
horizon degrees of freedom provide a holographic description of the
black hole together with low-energy excitations of the spacetime
fields around the black hole.  This shows that for a consistent
holographic description we must take the proper temperature at the
stretched horizon to be equal to the Planck temperature.

%%%%%%%%%%%%%%%%%%%%%%%%%%%%%%%%%%%%%%%%%%%%%%%%%%%%%%%%%%%%%%%%%%%%%%%%%%%%%%%%
\section{Discussion and conclusions}
%%%%%%%%%%%%%%%%%%%%%%%%%%%%%%%%%%%%%%%%%%%%%%%%%%%%%%%%%%%%%%%%%%%%%%%%%%%%%%%%

We have proposed a quasi-particle description of 0-brane quantum
mechanics, and shown that certain simple properties of the
quasi-particle gas are responsible for the entropy--area relationship.
These properties are just the energy $\sim T$ and lifetime $\sim 1/T$
of the individual quasi-particles.  The quasi-particle picture
correctly reproduces many properties of the black hole; this strongly
suggests that the quasi-particle picture is correct.  In appendix A we
give further evidence that the quasi-particle description can be
derived from a mean-field approximation to the quantum mechanics.

Let us make a few comments on our main result (\ref{SA}).  First, the
Stefan-Boltzmann argument depends crucially on putting the stretched
horizon at the Planck temperature.  This is necessary to have a
consistent holographic description, as we saw for example in section
\ref{thermalization}.  Nonetheless it is worthwhile examining other
motivations for this choice.  We want to use (\ref{therma}) to extract
the thermalization rate of the quantum mechanics.  For this purpose
one must get close enough to the horizon.  One might think that
staying one string length outside the horizon is the right thing, but
this is not the case.  The energy stored in closed string modes
outside a sphere one string length away from the horizon is of order
one in 't Hooft units, while the black hole mass is of order $N^2$.
Thus one cannot expect to pick up enough degrees of freedom if one
stops at the string length.  In contrast it is well known from brick
wall models \cite{thooft} that at one Planck length outside the event
horizon the energy stored in closed string modes becomes of order the
black hole mass.

Second, the use of the Stefan--Boltzmann law to describe radiation
from the stretched horizon is only an approximation.  The relation of
Stefan--Boltzmann to a more precise calculation is as follows.
Consider a scalar field in the Hartle--Hawking vacuum.  In terms of
Schwarzschild modes the Hartle--Hawking vacuum looks thermal, with an
outgoing flux of radiation from the horizon.  There is an energy flux
associated with this outgoing radiation.  For large black holes, where
the near-horizon geometry is approximately Rindler, the flux of energy
is given by the Stefan--Boltzmann law \cite{BD}.

We should point out similarities of our picture with previous ideas.
In a gauge-fixed formalism we identify the quasi-particles with the
entries in the matrix fields $X(t)$.\footnote{For a discussion of
gauge symmetry in the mean-field approximation see \cite{kll}.}  We
take these to be a convenient set of degrees of freedom to describe
the black hole. Of course gauge invariant operators are formed from
traces of products of the matrices $X(t)$, but such operators do not
form a convenient set. This reminds us of the picture that black hole
energy and entropy are encoded in open strings (or half closed
strings) stuck to the horizon of the black hole \cite{speculations}.
Our introduction of a stretched horizon is suggested by the membrane
paradigm.  Note that the timescale for linear response of a classical
membrane has been shown to be of order the inverse temperature
\cite{membrane}, in agreement with our results.  Finally, our choice
of a special distance just outside the event horizon is motivated by
the brick wall model \cite{thooft}.

\bigskip
\centerline{\bf Acknowledgements}
\noindent
We are grateful to Gary Horowitz, Samir Mathur and Lenny Susskind for
valuable discussions.  NI and DK are supported by the DOE under
contract DE-FG02-92ER40699.  The research of DL is supported in part
by DOE grant DE-FE0291ER40688-Task A.  DK, DL and GL are supported in
part by US--Israel Bi-national Science Foundation grant \#2000359.

%%%%%%%%%%%%%%%%%%%%%%%%%%%%%%%%%%%%%%%%%%%%%%%%%%%%%%%%%%%%%%%%%%%%%%%%%
\appendix
\section{Mean-field approximation}
%%%%%%%%%%%%%%%%%%%%%%%%%%%%%%%%%%%%%%%%%%%%%%%%%%%%%%%%%%%%%%%%%%%%%%%%%

In \cite{kl,kll} we developed and applied a mean-field approximation
scheme to the quantum mechanics of $N$ D0-branes at finite
temperature.  Working in the imaginary-time formalism, the
approximation gives a set of two-point correlators evaluated at
Matsubara frequencies.  In this section we will focus on a particular
scalar propagator, denoted $\Delta^2(k)$ in \cite{kll}.  In this
section we argue that the mean-field approximation gives rise to a
quasi-particle picture of the quantum mechanics which is compatible
with the properties that we have been discussing.

To extract quasi-particle properties from mean-field correlators, it
is useful to introduce a spectral representation for the correlators.
Thus we write the Euclidean propagator as
\be
\label{SpectralRep}
\Delta^2(k) = \int_0^\infty d\omega \, \rho(\omega) {1 \over k^2 + \omega^2}
\ee
where the spectral density is given by
\be
\label{SpectralDensity}
\rho(\omega) = {1 \over Z} \sum_m e^{-\beta E_m} \sum_{n > m} \big\vert
\langle n \vert \phi \vert m \rangle \big\vert^2 \, 2 \omega \left(1 - e^{-\beta \omega}\right)
\delta(\omega - E_n + E_m)\,.
\ee
One can regard $\rho(\omega) d\omega$ as the number of single-string
states with energy between $\omega$ and $\omega+d\omega$.  In
principle, the spectral density is uniquely determined by the
Euclidean propagator evaluated at the Matsubara frequencies
\[
k_l = 2 \pi l / \beta \qquad l \in {\mathbb Z}
\]
together with some information about behavior at infinity
\cite{BaymMermin}.  But in practice it is extremely
difficult to invert (\ref{SpectralRep}) to solve for $\rho(\omega)$.

Similar spectral representations were used in \cite{ikll}, where
mean-field methods were applied to study the dynamics of a 0-brane
probe of the black hole background.  In that work we analytically
continued the probe gap equations to determine the probe propagators
in between the Matsubara frequencies.  This additional information
made it possible to uniquely determine the spectral density of the
probe using standard inverse-problem methods.  A striking feature
found in \cite {ikll} was that, for small probe radius $R$, the
spectral density consists of two well-defined and well-separated
peaks.  Once $R$ becomes smaller than the radius corresponding to the
string-scale stretched horizon the probe becomes indistinguishable
from the 0-branes that make up the black hole.  We therefore expect
the spectral density for the black hole background to have two
well-defined peaks.

To determine the spectral density for the propagators describing the
black hole background we could proceed as in \cite{ikll}, and
analytically continue the gap equations of \cite{kll} away from the
Matsubara frequencies.  But rather than carry out this rather involved
procedure, we begin by applying inverse-problem methods to the
propagator evaluated only at the Matsubara frequencies.\footnote{We
determine the propagator by solving the gap equations given in (32) --
(40) of \cite{kll}.}  We apply the Tikhonov regularization method
described in \cite{ikll} to obtain the spectral density subject to a
positivity constraint. In figure 1 we show the results for the
spectral density at inverse temperature $\beta=2$.  A two-peak
structure of the density of states clearly emerges.  We also see that
the low-frequency peak is symmetrical, and extends down to close to
zero frequency.  This behavior seems to persist as we lower the
temperature, at least to $\beta = 4$.  Thus we take the width of the
low-frequency peak to be of order its mass.

\begin{figure}
\epsfig{file=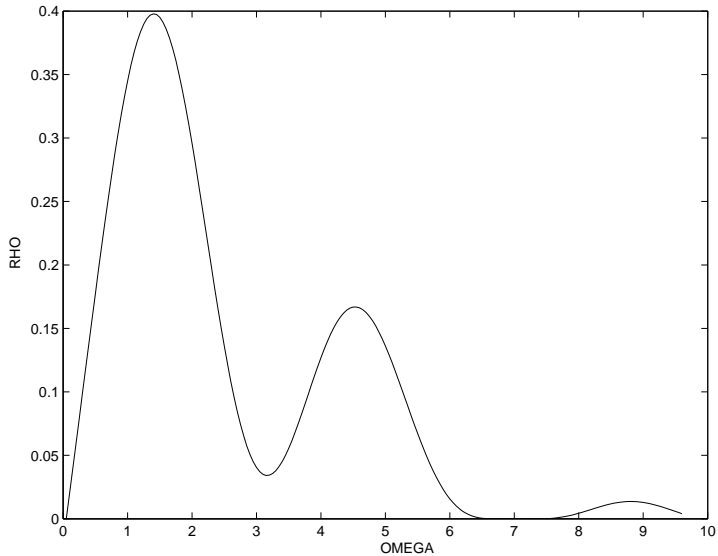,height=75mm}
\caption{Spectral density $\rho$ as a function of frequency $\omega$.}
\end{figure}

We should mention a number of caveats in this analysis. It is possible
that the peak width observed here is an artifact of evaluating the
propagator only at Matsubara frequencies. However the convergence of
the regularization method, and its independence of the frequency
resolution used, are signs that the density of states obtained this
way is reliable. For $\beta > 4$ the method does not give convergent
results.  A more precise determination of the widths will be presented
in future work on the continued gap equations.

To determine the dependence of the masses and widths on temperature,
we proceed to make an ansatz for the form of the propagator, and
determine the parameters appearing in the ansatz by fitting to the
propagator evaluated at Matsubara frequencies. This will allow us to
analyze inverse temperatures $\beta >4$.  Motivated by the results of
\cite{ikll}, we expect the retarded propagator to have two pairs of
poles in the lower half plane.
\[
G_R(k) = - {A_1 \over (k + i \Gamma_1/2)^2 - m_1^2} - {A_2 \over (k + i \Gamma_2/2)^2 - m_2^2}~.
\]
This corresponds to two species of quasi-particles, with masses $m_1$,
$m_2$ and lifetimes $1/\Gamma_1$, $1/\Gamma_2$.  The corresponding
spectral density is
\[
\rho(\omega) = {2\omega \over \pi} {\rm Im} \, G_R(\omega) = {\omega \over 2 \pi}
\sum_{i = 1,2} {A_i \Gamma_i \over m_i} \left({1 \over (\omega - m_i)^2 + \Gamma_i^2/4}
- {1 \over (\omega + m_i)^2 + \Gamma_i^2/4}\right)
\]
which in turn implies that our ansatz for the Euclidean Green's
function is
\be
\label{ansatz}
G_E(k) = {A_1 \over (\vert k \vert + \Gamma_1/2)^2 + m_1^2} +
{A_2 \over (\vert k \vert + \Gamma_2/2)^2 + m_2^2}\,.
\ee
Note that the Euclidean propagator is non-analytic at zero momentum.

The spectral density satisfies the sum rule $\int_0^\infty d\omega
\rho(\omega) = 1$ which implies that $A_1 + A_2 = 1$
\cite{FetterWalecka}.  This leaves five parameters to be determined.
Unfortunately, it does not seem possible to reliably extract the
values of all five parameters just given the propagators at Matsubara
frequencies.  We therefore fix the widths by hand, and minimize
\[
\chi^2 = \sum_{l=0}^\infty \left(\Delta^2(2 \pi l / \beta) - G_E(2 \pi l / \beta)\right)^2
\]
to determine the best-fit values of the remaining three parameters
$m_1$, $m_2$, $A_1 = 1 - A_2$.  Without loss of generality we take
$m_1 < m_2$.

\begin{figure}
\epsfig{file=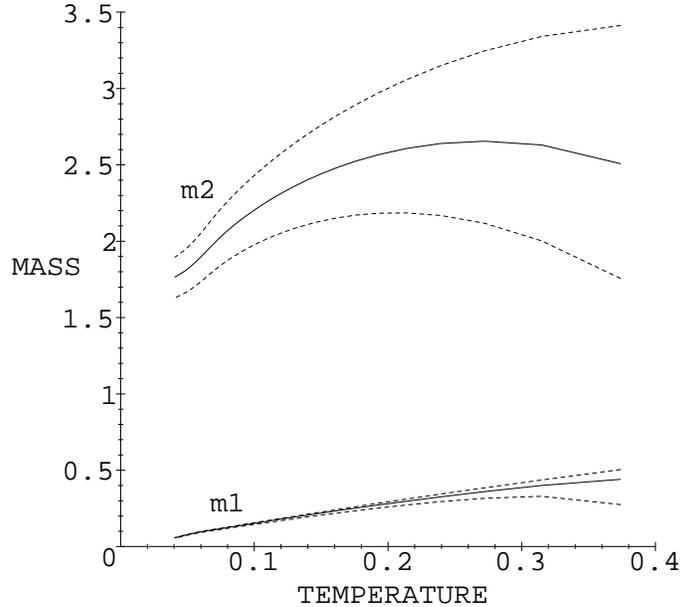,height=100mm}
\caption{Plots of $m_1$ and $m_2$ vs.~temperature, for various choices of widths.}
\end{figure}

Results for $m_1(T)$ and $m_2(T)$ are shown in Fig.~2, where we have
allowed $\Gamma_1$ and $\Gamma_2$ to vary over the ranges
\be
\label{WidthRange}
T/2 < \Gamma_1 < m_1 \qquad T/2 < \Gamma_2 < m_2/2\,.
\ee
We find that we get quite good fits over these ranges.  As can be seen
in the figure, the results for $m_1$ are not particularly sensitive to
our choice for the widths.  The three curves for $m_1$ shown in Fig.~2
are all well fit by $m_1 \approx 1.5 T$.  This linear dependence of
$m_1$ on temperature is consistent with the density of states
from Tikhonov regularization.  Likewise the range of widths $\Gamma_1$
compatible with a given $m_1$ are consistent with the regularization
results that indicate a linear relation between width and mass.

The area under the first peak in the spectral density $A_1$ does not
seem to be particularly sensitive to our choice for the widths.  At
very low temperatures $A_1$ falls off like a power law and is well fit
by $A_1 \approx 1.2 \beta^{-0.9}$.  $A_1$ should really fall off
faster with $\beta$ in order to reproduce the black hole entropy
$S_{\rm BH} \sim \beta^{-1.8}$.  This is a shortcoming of the
particular gap equations we have used, which only correctly reproduce
the entropy up to $\beta \sim 4$. This difficulty is discussed in more
detail in \cite{kll}.

For $\beta > 2.5$ the ansatz (\ref{ansatz}) provides a very good fit
to the propagators.  $\chi^2$ increases with $\beta$, but even at
$\beta = 25$ for widths in the range (\ref{WidthRange}) we have
\beas
&& \chi^2 \approx 7 \times 10^{-3} \\
\noalign{\vskip 2mm}
&& \max_l \Big\vert \Delta^2(2\pi l/\beta) - G_E(2\pi l/\beta)
\Big\vert \approx 0.05 \quad \hbox{\rm (off by about 10\%)}~.
\eeas
Given that the goodness-of-fit is quite insensitive to our choice of
the widths we cannot determine the widths with any accuracy.  At high
temperatures, roughly $\beta < 0.5$, the propagator is well fit by a
spectral density with a single sharp peak at a frequency $m \approx
1.8/\beta^{1/4}$.\footnote{This follows from the high-temperature
analysis in \cite{kll}.  In the notation of that paper we are setting
$m^2 \approx 1/\Delta_0^2 \approx m_\Delta^2$.}  The double Lorentzian
ansatz does not give a good fit to the propagators in the range $0.5 <
\beta < 2.5$. We expect that only the more general inverse-problem
methods will be useful in pinning down the spectral density in this
regime.

%%%%%%%%%%%%%%%%%%%%%%%%%%%%%%%%%%%%%%%%%%%%%%%%%%%%%%%%%%%%%%%%%%%%%
\section{Quasinormal modes}
\label{quasinormal}
%%%%%%%%%%%%%%%%%%%%%%%%%%%%%%%%%%%%%%%%%%%%%%%%%%%%%%%%%%%%%%%%%%%%%

In this appendix we compute the thermalization time of the black hole
by studying quasi-normal excitations of the black hole background.
Note that this computation is independent of the considerations
leading to (\ref{therma2}).

Quasinormal modes for black holes in asymptotically AdS spaces were
studied in \cite{hh, CardosoLemos, b, bss, SonStarinets}.  The
starting point is the near-horizon Einstein frame metric
\bea
\nonumber
ds^2 & = & {\rm const.} \, U^{21/8} \left[-h(U)dt^2 + h^{-1}(U)dU^2 + {c^{1/2}
(g_{YM}^2 N)^{1/2} \over U^{3/2}} d\Omega_{8}^{2}\right] \\
\label{metric}
h(U) & = & \frac{U^{7/2}}{c^{1/2} (g_{YM}^2 N)^{1/2}}\left(1-\frac{U_{0}^{7}}{U^{7}}\right)\,.
\eea
Here $c = 240 \pi^5$ and $g^2_{YM}$ is the coupling constant of the
dual gauge theory.  The horizon is located at $U = U_0$, while the
`boundary of space' is at $U = \infty$.  The dilaton, for example,
obeys the minimal scalar wave equation $\nabla^2 \phi = 0$.
Separating variables $\phi(t,U,\Omega) = e^{-i \omega t} \phi(U)
Y_\ell(\Omega)$ where $Y_\ell$ is a spherical harmonic on $S^8$ leads
to the radial wave equation
\[
\partial_U\left(U^8\Big(1 - {U_0^7 \over U^7}\Big)\partial_U\phi\right)
+ \left({c g_{YM}^2 N \omega^2 U \over 1 - U_0^7/U^7} - \ell(\ell+7)U^6\right)\phi = 0\,.
\]
Introduce a new dimensionless radial coordinate $x = - \log\left(1 -
U_0^7/U^7\right)$.  The horizon is at $x = \infty$, while the boundary
of space is at $x=0$.  The wave equation takes the canonical form
\[
\left({d^2 \over dx^2} + {\rho^2 \over (1 - e^{-x})^{9/7}}
- {\ell(\ell+7) \over 49} {e^{-x} \over (1 - e^{-x})^2}\right)\phi = 0
\]
where the dimensionless parameter $\rho$ is
\[
\rho = {c^{1/2} (g^2_{YM} N)^{1/2} \omega \over 7 \, U_0^{5/2}}\,.
\]
Quasinormal frequencies are determined by requiring ingoing waves at
the future horizon, while as in \cite{hh} we will impose Dirichlet
boundary conditions at $x=0$.
\bea
\nonumber
\phi(t,x) & \sim & e^{i \rho x - i \omega t} \qquad \hbox{\rm as $x\to
\infty$ \ {\rm with} \ $x - \omega t/\rho$ \ {\rm fixed}} \\
\label{bcs}
\phi &=& 0 \qquad\qquad\,\,\, \hbox{\rm at $x = 0$}~.
\eea
The main point now follows simply from dimensional analysis.  The
boundary conditions (\ref{bcs}) can only be satisfied for discrete
complex values of $\rho$, labeled by a radial quantum number $n$ and
an angular quantum number $\ell$.  Thus
\[
\rho_{n\ell} = f(n,\ell)
\]
or equivalently
\[
\omega_{n\ell} \sim {U_0^{5/2} \over (g^2_{YM} N)^{1/2}} f(n,\ell) \sim T f(n,\ell)\,.
\]
This shows that both the real and imaginary parts of the quasinormal
frequencies are proportional to the Hawking temperature of the black
hole.  In the dual picture this means that the thermalization time of
the quantum mechanics is proportional to $1/T$.  A similar scaling
argument for conformal $p$-branes was presented in \cite{hh}.  Related
scaling arguments work for the near horizon geometry of all
$p$-branes, and show that the thermalization time for all these
theories is proportional to $1/T$ \cite{tc}.

Determining the quasinormal frequencies requires some numerical
analysis.  It is convenient to work in terms of a new radial variable
$w = e^{-x}$, which puts the horizon at $w=0$ and the boundary of
space at $w=1$.  For s-waves the radial wave equation becomes
\[
\left(w {d \over dw} w {d \over dw} + {\rho^2 \over (1 - w)^{9/7}} \right) \phi = 0\,.
\]
This equation can be solved in a power series about $w=0$.  The
solution which is ingoing at the horizon has the expansion
\[
\phi(w) = w^{-i\rho}\left(1 - {9 \rho^2 \over 7 - 14 i \rho} w + \cdots \right)\,.
\]
As in \cite{hh} we truncate the series at some finite order $N$;
imposing the Dirichlet condition $\phi\vert_{w = 1} = 0$ then gives an
algebraic equation for $\rho$.  We have carried out this procedure
keeping terms up to order $w^{19}$, which gives the lowest quasinormal
frequency
\[
\rho = 0.56 - 0.93 i\,.
\]
This result is fairly stable; the lowest quasinormal frequency does
not change significantly with the truncation order provided $N > 10$.

These quasi-normal frequencies will show up as poles in the complex
frequency plane of correlators of supergravity excitations
\cite{bss}. We may map such correlators into correlation functions of
scaling operators in the quantum mechanics, following the dictionary
worked out in \cite{yoneya}. Such correlators reduce to convolutions
of products of the elementary quasi-particle two-point functions. The
width of the first peak in the quasi-particle spectral density should
therefore exhibit the same temperature dependence as the quasi-normal
mode frequencies.

%%%%%%%%%%%%%%%%%%%%%%%%%%%%%%%%%%%%%%%%%%%%%%%%%%%%%%%%%%%%%%%%%%%%%%%%%%%%%%%%

\end{document}